\documentclass[12pt]{iopart}
\usepackage{iopams}
\usepackage{graphicx}

\begin{document}

\title[Work distribution in logarithmic-harmonic potential] {Work
  distribution in time-dependent logarithmic-harmonic potential: exact
  results and asymptotic analysis}

\author{Artem Ryabov$^{1}$, Marcel Dierl$^{2}$, Petr Chvosta$^{1}$,
  Mario Einax$^{2}$, Philipp Maass$^{2}$}

\address{$^{1}$ Department of Macromolecular Physics, Faculty of
  Mathematics and Physics, Charles University in Prague, V Hole{\v
    s}ovi{\v c}k{\' a}ch 2, CZ-180~00~Praha, Czech Republic}

\address{$^{2}$ Fachbereich Physik, Universit\"at Osnabr\"{u}ck,
  Barbarastra{\ss}e 7, 49076 Osnabr\"{u}ck, Germany}
\eads{\mailto{rjabov.a@gmail.com}}

\date{13 December, 2012}


\begin{abstract} 
  We investigate the distribution of work performed on a Brownian
  particle in a time-dependent asymmetric potential well. The
  potential has a harmonic component with time-dependent force
  constant and a time-independent logarithmic barrier at the origin.
  For arbitrary driving protocol, the problem of solving the
  Fokker-Planck equation for the joint probability density of work and
  particle position is reduced to the solution of the Riccati
  differential equation. For a particular choice of the driving
  protocol, an exact solution of the Riccati equation is
  presented. Asymptotic analysis of the resulting expression yields
  the tail behavior of the work distribution for small and large work
  values. In the limit of vanishing logarithmic barrier, the work
  distribution for the breathing parabola model is obtained.
\end{abstract}

\maketitle

\section{Introduction}

Stochastic thermodynamics is an advancing field with many applications
to small systems of current interest \cite{Seifert2008, EspositoI,
  EspositoII}.  Work performed on a small system by an external
driving becomes a stochastic variable because of the strong influence
of fluctuations mediated by the environment. In a standard setting, a
particle moves in a thermal environment and experiences a
time-dependent external force. The particle position by itself is a
stochastic process, say $\mathsf{X}(t)$. Any single trajectory of the
particle in a time interval $[0,t]$ yields a single value of the work
$\mathsf{W}(t)$ done on the particle. The work $\mathsf{W}(t)$ thus is
a functional of the position process $\mathsf{X}(t')$, $0\le t'\le t$,
and is distributed with a probability density function (PDF)
$p(w;t)$. The probability $p(w;t){\rm d}w$ that the work
$\mathsf{W}(t)$ falls into an infinitesimal interval $(w,w+{\rm d}w)$
equals the probabilistic weight of all trajectories giving work values
in that interval.  An important aspect of stochastic thermodynamics is
that relations between work distributions for forward and reversed
driving protocols as well as averages over functions of the work allow
one to determine thermodynamic potential differences between
equilibrium states of small systems. Perhaps most widely known is the
Jarzynski equality \cite{Jarzynski:1999}, which relates the free
energy difference between two equilibrium configurations of a system
to the average
\begin{equation}
\label{averaging}
\left\langle {\rm e}^{-\beta \mathsf{W}(t)} \right\rangle =
\int_{-\infty}^{+\infty}\!\!\!{\rm d}w\,
{\rm e}^{-\beta w}p(w;t)\,\,,
\end{equation}
where $\beta^{-1}=k_{{\rm B}}T$ is the thermal energy. In computer
simulations and experiments, sampled values of $w$ lie typically
within one or two standard deviations of the maximum of $p(w;t)$,
while values most important in Eq.~(\ref{averaging}) are those near
the maximum of $\exp(-\beta w)\,p(w;t)$. The corresponding regimes
overlap significantly only if the function $\exp(-\beta w)$ does not
change much over one standard deviation. If this is not the case, the
relevant contributions to the integral come from those rare
trajectories with work value belonging to the tails of $p(w;t)$.  In
experiments these rare trajectories are almost never observed and even
in simulations it is difficult to generate them with sufficient
statistical weight. Hence, to evaluate averages as in
Eq.~(\ref{averaging}), information on the tails of the PDF is
essential. An important part of this study is to gain insight into the
asymptotic PDF behavior of the work performed on a Brownian particle
in an asymmetric time-dependent potential well.

We will calculate the characteristic function for the work in a simple
setting which, however, may be realized in experiments \cite{Kessler,
  Cohen,Blickle}. In this setting, an overdamped motion of a Brownian
particle is considered in the logarithmic-harmonic potential
\begin{equation}
\label{potential}
U(x,t) = -g \ln (x) + \frac{1}{2}k(t) x^{2}\,,\quad
g>0\,,
\quad
x>0\,.
\end{equation}
where the parameter $g>0$ specifies the strength of the logarithmic
repulsive wall near the origin, and $k(t)$ is a time-dependent force
constant. In the deterministic limit, i.e., in absence of thermal
noice, the particle moves along the positive $x$-axis as driven by the
time-dependent force $-\partial U(x,t)/\partial x$ (without inertia).
Taking into account the thermal noise, the combined process
$\{\mathsf{X}(t),\mathsf{W}(t)\}$ is described by the system of
Langevin equations
\begin{equation}
\label{Langevin1}
{\rm d}\mathsf{X}(t)=
\left[\frac{g}{\mathsf{X}(t)}-k(t)\mathsf{X}(t)\right]{\rm
  d}t+\sqrt{2D}\,{\rm d}\mathsf{B}(t)\,,
\end{equation}
\begin{equation}
\label{Langevin2}
{\rm d}\mathsf{W}(t)=
\frac{1}{2}\dot{k}(t)\mathsf{X}^{2}(t)\,{\rm d}t\,,
\end{equation}
where $D$ quantifies the strength of the noise and $\mathsf{B}(t)$ is
the standard Wiener process.  Specifically, in the case of a thermal
environment, the noise strength is proportional to the temperature,
$D=k_{{\rm B}}T$ (when the particle mobility is set to one).

While the work process has not yet been studied for the
logarithmic-harmonic potential, an exact solution of the Fokker-Planck
equation for the position process was obtained \cite{Giampaoli}.  In
the following, we recover this solution from our Lie algebraic
approach.  The solvability of this problem stems from the fact that
the operators entering the Fokker-Planck equation form a Lie algebra
\cite{WeiNorman, Wolf}.  If one considers the Fokker-Planck equation
for the joint PDF of position and work, the corresponding differential
operators no longer form a Lie algebra.  However, if one starts with
the Fokker-Planck equation for the joint PDF and performs a Laplace
transformation with respect to the work variable $w$, a Lie algebra is
obtained. Solution of the Fokker-Planck equation then provides the
characteristic function for the work process and the tails of the work
PDF can be extracted using asymptotic analysis of Laplace transforms.

The work PDF for the problem of the ``breathing parabola'' ($g=0$) has
been studied analytically in Refs.~\cite{Engel},
\cite{NickelsenEngel}, \cite{Speck}, and \cite{MinhAdib}. In
\cite{Engel}, \cite{NickelsenEngel}, the authors considered an
expansion around a single trajectory attributed to a prescribed rare
value of the work and derived asymptotic results for the tails of the
work PDF in the small temperature limit. The solution reported in
\cite{Speck} is formally exact for arbitrary protocol $k(t)$. Explicit
results are given in the limit of slow driving, where the process is
close to a quasi-static equilibrium and the work PDF can be
approximated by a Gaussian. In Ref.~\cite{MinhAdib} the work-weighted
propagator was derived by the path integral method. Another closely
related setting, where the work PDF can be calculated analytically, is
for a parabolic potential with a time-dependent position of the
minimum (``sliding parabola'') \cite{MinhAdib},
\cite{MazonkaJarzynski}, \cite{ZonCohenI}, \cite{ZonCohenII},
\cite{EDGCohen}.  The present work broadens the list of few exact
results in this field.

\section{Solution of Fokker-Planck equation for arbitrary protocol}
\subsection{Green's function for logarithmic potential}
\label{Subsection2.1}
An important auxiliary quantity in deriving all subsequent results
is the Green's function
\begin{equation}
\label{J0}
q(x;t| x_{0}) = \exp\!\!\left[\,t\! \left( \frac{\partial^2 }{\partial x^{2}}
- \frac{g}{D} \frac{\partial }{\partial x} \frac{1}{x} \right) \right]
\delta(x\! -\! x_{0})\,, \quad x_{0}>0\,,
\end{equation}
which represents the solution of the Fokker-Planck equation
\begin{equation}
\frac{\partial }{\partial t} q(x;t| x_{0}) = \left(
 \frac{ \partial^2 }{\partial x^{2}}
- \frac{g}{D} \frac{\partial }{\partial x} \frac{1}{x} \right) q(x;t| x_{0})\,\,
\end{equation}
for the diffusion in the time-independent logarithmic potential with
initial condition $q(x;0|x_{0})=\delta(x\!-\! x_{0})$. The explicit
form of the solution is \cite{KarlinTaylor,Bray}
\begin{equation}
q(x;t| x_{0}) = 
\frac{ x_{0}}{2t} \left( \frac{x}{x_{0}} \right)^{\!\! \nu + 1} \!\!\!\!\!
	\exp\!\!\left(\!-\frac{x^{2}+x_{0}^{2}}{4t}\right)
	 I_{\nu}\!\!\left(\frac{x x_{0}}{2t}\right)\,\,,	
 \label{qdef}
\end{equation}
where $I_\nu(.)$ is the modified Bessel function of order $\nu$, and
\begin{equation}
\nu = \frac{1}{2}\left( \frac{g}{D} -1 \right)
\label{nudef}
\end{equation}
measures the strength of the logarithmic potential in relation to the
intensity of the thermal noise. In all subsequent results the
parameter $g$ enters solely through $\nu$ defined in Eq.~(\ref{nudef}).

Equation~(\ref{qdef}) is the unique norm-preserving solution of the
diffusion problem in the domain $x>0$, i.e., the probability current
at $x=0$ vanishes. Therefore, performing the limit $g \to 0$ and using
$I_{-1/2}(z)=\sqrt{2/\pi}\cosh(z)/\sqrt{z}$, we obtain the standard
solution for free diffusion with a {\em reflecting boundary\/}
at $x=0$ \cite{Redner}.

\subsection{Joint Green's function for work and position}
\label{Subsection2.2}
Let us denote by $p(x,w;t|x_{0},0)$ the joint PDF for the process
$\{\mathsf{X}(t),\mathsf{W}(t)\}$ given that at time $t=0$ the
particle is at position $x_{0}$, $x_{0}>0$, and no work has been done
yet,
\begin{equation}
p(x,w;0|x_{0},0) = \delta(x\! - \! x_{0}) \delta(w)\,, 
\qquad x_{0}>0\,\,.
\end{equation}
The time evolution of the joint PDF is given by the Fokker-Planck
equation
\begin{equation}
  \fl
  \frac{\partial }{\partial t} p(x,w;t|x_{0},0) = 
  \left[ D \frac{\partial^2 }{\partial x^2} -
    \frac{\partial }{\partial x} \left(\frac{g}{x}-k(t)x\right) 
    -  \frac{ 1}{2}\dot{k}(t)x^{2} \frac{\partial}{\partial w} \right]
  p(x,w;t|x_{0},0)\,.
\label{FPoperator}
\end{equation}
The differential operators on the right-hand side \emph{do not} exhibit
closed commutation relations. However, after performing the two-sided
Laplace transformation \cite{Doetsch}
\begin{equation}
\label{Laplace}
 \widetilde{p}(x,\xi ;t|x_{0}) = \int_{-\infty}^{+\infty}\!\!\!\! {\rm d}w \,
 {\rm e}^{- \xi w} p(x, w;t|x_{0},0)\,\,,
\end{equation}
the Fokker-Planck equation (\ref{FPoperator}) assumes the form
\begin{equation}
\label{xiFPoperator}
\fl
\frac{\partial }{\partial t} 
\widetilde{p}(x,\xi;t|x_{0},0)=
\left[D \widehat{J}_{0} + 2 k(t) \widehat{J}_{1} - 
4 \xi \dot{k}(t) \widehat{J}_{2} + \left(\nu + 1\right)k(t)\right]
\widetilde{p}(x,\xi;t|x_{0},0)\,\,,
\end{equation}
where the differential operators
\begin{equation}
\widehat{J}_{0} = \frac{\partial^2 }{\partial x^2} -
\frac{g }{D} \frac{\partial }{\partial x}\frac{1}{x} \,\,,
\qquad
\widehat{J}_{1} = \frac{1}{2} \left( x \frac{\partial }{\partial x}  
-\nu  \right) \,\,,
\qquad
\widehat{J}_{2} = \frac{1}{8} x^2\,\,,
\end{equation}
satisfy the \emph{closed} commutation relations
\begin{equation}
[\widehat{J}_{0} , \widehat{J}_{1}] = \widehat{J}_{0} \,\,, \qquad
[\widehat{J}_{0} , \widehat{J}_{2}] = \widehat{J}_{1} \,\,, \qquad
[\widehat{J}_{1} , \widehat{J}_{2}] = \widehat{J}_{2} \,\,.
\end{equation}
This allows us to apply the Lie algebraic method, as discussed, e.g.,
in Refs.~\cite{WeiNorman,Wilcox}. First, we write the solution of
Eq.~(\ref{xiFPoperator}) in the factorized form
\begin{eqnarray}
\label{factorizedsolution}
\fl
\widetilde{p}(x,\xi ;t|x_{0}) &=&
\exp\!\!\left[ (\nu\! +\! 1)\!\! \int_{0}^{t}\!\!\! {\rm d}t' k(t') \right]\!
\exp\!\!\left( b_{2}(t) \widehat{J}_{2} \right)
\exp\!\!\left( b_{1}(t) \widehat{J}_{1} \right)
\exp\!\!\left( b_{0}(t) \widehat{J}_{0} \right)\! \delta(x \!-\!
x_{0})
\end{eqnarray}
where the time-dependent coefficient $b_{2}(t)$ is obtained by solving
the Riccati differential equation
\begin{equation}
\label{Riccati}
\dot{b}_{2}(t) = \frac{D}{2}b_{2}^{2}(t) + 2 k(t) b_{2}(t)- 4 \xi \dot{k}(t)\,,
\qquad  b_{2}(0)=0\,.
\end{equation}
Knowing $b_2(t)$ the other coefficients are given by
\begin{equation}
\label{b0b1}
b_{1}(t) = 2 \int_{0}^{t}\!\! {\rm d}t'\, k(t') 
+ D\! \int_{0}^{t}\!\! {\rm d}t'\, b_{2}(t')\,,\quad
b_{0}(t) = D\! \int_{0}^{t}\!\! {\rm d}t' 
\exp\!\left[\, b_{1}(t')\, \right]\,.
\end{equation}
In the last step, we evaluate, using Eqs.~(\ref{J0}) and (\ref{qdef}),
the action of the operator $\exp(b_{0}(t)\widehat{J}_{0})$ on the
delta function in Eq.~(\ref{factorizedsolution}), and subsequently
apply to the corresponding result the two remaining exponential
operators in Eq.~(\ref{factorizedsolution}). This yields
\begin{eqnarray}
\label{finalresult}
\fl
\widetilde{p}(x,\xi ;t|x_{0}) =
\exp\!\!\left(  \int_{0}^{t}\!\!\! {\rm d}t' k(t')
\! -\! \frac{1}{2}\,\nu D\!\! \int_{0}^{t}\!\!\! {\rm d}t' b_{2}(t')\!
+\! \frac{1}{8}\, b_{2}(t)x^{2} \!\right)\!
q(x{\rm e}^{\frac{1}{2}\,b_{1}(t)}; b_{0}(t)|x_{0}) \,\,.
\end{eqnarray}
In the derivation we have utilized the operator identity
\begin{equation}
  \exp\left(\eta x \frac{\partial}{\partial x}\right)f(x)=
f\left(x\exp(\eta)\right)\,\,.
\end{equation}
The exact solution (\ref{finalresult}) is the central result of the
present section and constitutes the starting point of all
subsequent analyses.

\section{PDF of particle position and its long-time asymptotics}

After integrating the joint PDF $p(x,w;t|x_{0},0)$ over the work
variable, the transition PDF $p(x;t|x_{0})$ for the particle
coordinate is obtained.  Equivalently, the $w$-integration is
accomplished by evaluating the result (\ref{finalresult}) at $\xi=0$
[cf.\ Eq.~(\ref{Laplace})]. Notice that the variable $\xi$ enters the
solution (\ref{finalresult}) only through the Riccati equation
(\ref{Riccati}). When taking $\xi=0$, this equation reduces to the
Bernoulli differential equation, where the unique solution satisfying
$b_{2}(0)=0$ is the trivial one, $b_{2}(t)=0$. The remaining
coefficients in Eq.~(\ref{finalresult}) are then given by
\begin{eqnarray}
  b_{1}(t) = 2 \int_{0}^{t}\!\! {\rm d}t'\, k(t')\,,\quad
  b_{0}(t) = D\! \int_{0}^{t}\!\! {\rm d}t' \exp\!\left[\, 
2 \int_{0}^{t'}\!\! {\rm d}t''\, k(t'')\, \right]\,.
\end{eqnarray}
Hence the PDF for the particle position reads\footnote{This result
  agrees with Eq.~(19) in Ref.\ \cite{Lo}, where it has been derived
  in connection with a diffusion problem with logarithmic factors in
  drift and diffusion coefficients.}
\begin{equation}
\label{PDFx}
p(x;t|x_{0})=\frac{x_{0}\,{\rm e}^{\frac{\nu+2}{2}\,b_{1}(t)}}{2b_{0}(t)}
\left( \frac{x}{x_{0}} \right)^{\!\! \nu + 1} \!\!\!\!\!
\exp\!\!\left(\!-\frac{x^{2}{\rm e}^{b_{1}(t)}\!+\! x_{0}^{2}}{4 b_{0}(t)}\right)
I_{\nu}\!\!\left(  \frac{ x x_{0}\, {\rm e}^{\frac{1}{2}\,b_{1}(t)}}{2 b_{0}(t)}
\right)\,.
\end{equation}
This finding is valid for an arbitrary driving protocol $k(t)$.  If
$k(t)$ is a positive constant, say $k(t)\!=\!k_{0}>0$, then the system
approaches the Gibbs canonical equilibrium at long times. If the
constant force constant is superimposed with a periodically
oscillating component, a gradual formation of a nontrivial steady
state occurs. In this steady state, the PDF does not depend on the
initial condition $x_{0}$ and, for any given $x>0$ is a periodic
function of time with the fundamental period given by that of $k(t)$.

To exemplify the PDF in the steady state, let us take
\begin{equation}
\label{nonmonotonous}
k(t) = k_{0}+ k_{1} \sin(\omega t)\,\,,\quad k_{0}>0\,.
\end{equation}
The asymptotic analysis of Eq.~(\ref{PDFx})
for long times requires the evaluation of the limit
\begin{equation}
  \lim_{t\to \infty} \frac{{\rm e}^{\alpha\,
      b_{1}(t)}}{b_{0}(t)}\,,
  \quad\alpha\leq 1\,\,.
\end{equation}
If $\alpha<1$, the limit exists and, using L'H{\^ o}pital's rule, it
equals zero.  Hence for any finite $x$ and $x_{0}$, the argument
$z=xx_0e^{b_1(t)/2}/(2b_0(t))$ in the Bessel function appearing in
Eq.~(\ref{PDFx}) becomes small for large $t$ and we can write
$I_{\nu}(z) \sim (\frac{1}{2}z)^{\nu}/\Gamma(\nu +1)$.  If
$\alpha=1$, the limit does not exist and ${\rm
  e}^{b_{1}(t)}/b_{0}(t)\sim 1/f(t)$ for $t\to\infty$, where
\begin{equation}
f(t) = D \exp\!\!\left(\frac{2k_{1}}{\omega}\cos(\omega t)\right)
\sum_{n=-\infty}^{+\infty}
I_{n}\!\!\left(\!-\frac{2k_{1}}{\omega}\!\right)
\frac{{\rm e}^{in\omega t}}{2k_{0}+i n \omega} \,\,, \quad k_{0}>0\,\,.
\end{equation}
Accordingly, for $t\to\infty$
\begin{equation}
p(x;t|x_0)\sim p_{\rm as}(x;t) =
\frac{1}{\Gamma(\nu +1)}
\left(  \frac{1}{f(t)} \right)^{\!\! \nu+1}
\!\!\!
\left(\frac{x}{2}\right)^{\! 2\nu + 1}
\!\!\!
\exp\!\!\left[- \left(\frac{x}{2}\right)^{\! 2} \!
 \frac{1}{f(t)} \right]\,.
\label{PDFxasy}
\end{equation}
In the limit $k_{1}\to 0$ or $\omega\to 0$, $f(t)\to D/(2k_{0})$, and
$p(x;t|x_0)$ approaches the Gibbs equilibrium distribution.

Finally, the limit $g\to 0$ in Eq.~(\ref{PDFx}) [Eq.~(\ref{PDFxasy})]
yields the exact transition PDF (exact time-asymptotic PDF) for
the breathing parabola model with reflecting boundary at the
origin. Since the parameter $g$ enters only via $\nu$ in
Eq.~(\ref{nudef}), $g\to 0$ corresponds to $\nu\to-\frac{1}{2}$ in
Eqs.~(\ref{PDFx}) and (\ref{PDFxasy}).

\section{Work fluctuations}
\subsection{Characteristic functions}
By integration of the joint PDF in Eq.~(\ref{finalresult}) over the
spatial variable $x$, we obtain the characteristic function for the
work done on the particle during the time interval $[0,t]$.  

Let us first consider the particle dynamics conditioned on the initial
position $x_{0}$. In this case the characteristic function for the
work reads
\begin{equation}
\Phi(\xi ;t| x_{0}) =
\int_{0}^{+\infty}\!\!\!{\rm d}x\, \widetilde{p}(x,\xi ;t | x_{0})\,\,.
\end{equation}
Carrying out the integration 
we find
\begin{equation}
\label{CF1}
\Phi(\xi ;t| x_{0}) =
\left( \frac{2{\rm e}^{b_{1}(t) - \frac{1}{2}D\! \int_{0}^{t}\!\!{\rm d}t' b_{2}(t')}}{2{\rm e}^{b_{1}(t)}-b_{0}(t)b_{2}(t)} \right)^{\!\! \nu +1}
\!\!\!\!\!
\exp\!\!\left[ \left(\frac{x_{0}}{2}\right)^{\! 2} \!
\frac{b_{2}(t)}{2{\rm e}^{b_{1}(t)}-b_{0}(t)b_{2}(t)}\, \right]\,\,.
\end{equation}

A physically more important situation is, when the particle coordinate
is initially equilibrated with respect to the initial value
$k(0)=k_{0}>0$ of the force constant. In order to obtain the
characteristic function for this situation we have to integrate over
$x_{0}$ the product of the characteristic function $\Phi(\xi ;t|
x_{0})$ and the equilibrium PDF $p_{\rm as}$ for $f(t)=D/(2k_{0})$
given in Eq.\ (\ref{PDFxasy}). The result is
\begin{equation}
\label{CF2}
\Phi(\xi;t) = \left\langle {\rm e}^{- \xi \mathsf{W}(t)} \right\rangle 
=\left( \frac{4k_{0}{\rm e}^{b_{1}(t) - 
\frac{1}{2}D\! \int_{0}^{t}\!\!{\rm d}t' b_{2}(t')}}
  {4k_{0}{\rm e}^{b_{1}(t)}-
\left(D+ 2k_{0}b_{0}(t)\right)b_{2}(t)} \right)^{\!\! \nu +1}\,\,.
\end{equation}
Note that for $\xi\!=\!\beta$, the average in Eq.~(\ref{averaging})
equals $\Phi(\beta;t)$. Equation~(\ref{CF2}) is valid for an arbitrary
driving protocol $k(t)$. The Laplace variable $\xi$ enters
$\Phi(\xi;t)$ implicitly through the function $b_{2}(t)$ [cf.\ the
Riccati equation (\ref{Riccati})] and through the functions $b_{1}(t)$
and $b_{0}(t)$ [cf.\ Eq.~(\ref{b0b1})].

In the limit $g\to 0$ ($\nu\to-\frac{1}{2}$),
Eqs.~(\ref{CF1},\ref{CF2}) give the corresponding characteristic
functions for the breathing parabola model with reflecting boundary at
$x\!=\!0$.  These characteristic functions are also valid for the
breathing parabola model {\em without\/} reflecting boundary, if
rather obvious changes are made of the meaning of the initial
coordinate $x_{0}$ in Eq.\ (\ref{CF1}), and of the initial Gibbs
equilibrium state underlying Eq.\ (\ref{CF2}).  In the breathing
parabola model without reflecting boundary, Eq.~(\ref{CF1}) is valid
for $x_{0}\in(-\infty,+\infty)$ and Eq.~(\ref{CF2}) corresponds to the
initial Gibbs equilibrium in the parabolic potential $U(x_{0}) =
k_{0}x_{0}^{2}/2$. The equivalence of the characteristic functions for
the problems with and without reflecting boundary is due to the symmetry
of the parabolic potential, which implies that {\em the work done on
  the particle that crosses the origin is the same as the work done on
  the particle reflected at the origin.\/} This reasoning can be
supported by an independent calculation if one notices that both
models, the present model with logarithmic-harmonic potential and the
breathing parabola one, have the same operator algebra.

\subsection{Simple example}

A driving protocol, where an explicit solution of the Riccati equation
(\ref{Riccati}) in terms of elementary functions can be given, is
\begin{equation}
\label{protocol1}
k(t) = \frac{k_{0}}{1+\gamma t}\quad,\quad k_{0}>0\quad,\quad\gamma>0\,\,.
\end{equation}
Notice that the same protocol was considered before in
Refs.~\cite{Engel, NickelsenEngel}. According to
Eq.~(\ref{protocol1}), the potential well widens with time and hence
the work done on the particle is negative for any $x>0$. 
The solution of the Riccati equation (\ref{Riccati}) reads
\begin{equation}
\label{b2exact1}
\fl
b_{2}(t)=-\frac{2}{D}\frac{\rm d}{{\rm d} t}
\ln\left\{
(1+\gamma t)^{ \frac{1}{2\gamma}\left[(2k_{0} +\gamma) - A(\xi)\right]}
\left[ (1+\gamma t)^{A(\xi)} -
\frac{(2k_{0} +\gamma)+A(\xi)}{(2k_{0} +\gamma)-A(\xi)}
\right]\right\}\,\,,
\end{equation}
where
\begin{equation}
A(\xi) = \sqrt{(2 k_{0} + \gamma)^2 - 8 k_{0} \gamma D \xi}\,\,.
\end{equation}
For simplicity, we take $\gamma=1$ and $k_{0}=1$ in the following. After
calculating $b_{0}(t)$ and $b_{1}(t)$ from Eqs.~(\ref{b0b1}),
the characteristic function in Eq.~(\ref{CF2}) becomes
\begin{equation}
\label{charfraceq}
\Phi(\xi;t)=
\left(
  \frac{2 A(\xi) (1+t)^{\frac{1}{2} \left[3+A(\xi)\right]}}
  { A(\xi) \left[(1+t)^{A(\xi) }+1\right] + (3-2 D \xi ) 
\left[(1+t)^{A(\xi)} -1 \right]}\right)^{\nu +1}.
\end{equation}
For $\xi=1/D$ in particular, we obtain $\Phi(1/D;t)=(1+t)^{\nu+1}$
which exemplifies the Jarzynski equality for the driving protocol
(\ref{protocol1}).

Successive derivatives of the characteristic function with respect
to $\xi$ evaluated at $\xi=0$ yield the cumulants of the work
distribution. The mean work done on the particle during the
time interval $[0,t]$ is
\begin{equation}
\left\langle\mathsf{W}(t) \right\rangle = - (\nu\! +\! 1) \frac{D}{9}
\left[ 6 \ln (1\!+\!t) + \frac{t^{3}+3t^{2}+3t}{(1+t)^3} \right] \,\,.
\end{equation}
It is a monotonically decreasing function of $t$, where for small
times, the decrease is linear, while in the long-time limit, it is
logarithmic. For the variance we find
\begin{equation}
\fl
\left\langle\mathsf{W}^{2}(t) \right\rangle - 
\left\langle\mathsf{W}(t)\right\rangle^2  
= (\nu+1)\! \left( \frac{D}{9}\right)^{\! 2}
\! \frac{t^{3}+3t^{2}+3t}{ (1+t)^6}\,
\left[t^{3}+3t^{2}+3t+24 (1\!+\!t)^3 \ln(1\!+\!t) \right]\,.
\end{equation}
This increases monotonically, where the increase is quadratic for
small times and logarithmic for long times.  The strength $g$ of the
logarithmic potential barrier enters the above formulae only through
the multiplicative prefactor $(\nu+1)$. This holds true for all
cumulants. For stronger repulsion, the particle predominantly diffuses
in a region further away from the origin. The decrease of its typical
potential energy results in a larger absolute value of the mean work. At
the same time, the width of the work PDF increases, since the initial
particle position is sampled from a broader Gibbs distribution.

Equation~(\ref{charfraceq}) entails the complete information about the
work distribution $p(w;t)$. In particular it allows one to derive the
tails of the work PDF for both $w\to 0^{-}$ and $w\to-\infty$ without
carrying out the inverse Laplace transformation of $\Phi(\xi;t)$.

The asymptotics of $p(w;t)$ for $w\to 0^{-}$ 
is related to the asymptotics of $\Phi(\xi;t)$ for $\xi \to-\infty$,
which follows from (\ref{charfraceq}),
\begin{equation}
  \Phi(\xi;t) \sim \left(\sqrt{8}\left(1+t\right)^{\frac{3}{2}}
    \frac{\exp\!\!\left[-\sqrt{2}\,\ln\!\left(1+t\right) 
\sqrt{-D\xi} \right] }{\sqrt{-D\xi}}
  \right)^{\nu +1 },
  \quad \xi \to - \infty\,.
\end{equation}
By taking the inverse Laplace transform of this asymptotic form
(cf.\ Ref.~\cite{Bateman}), we obtain the parabolic cylinder function
(cf.\ Ref.~\cite{Erdelyi}) with argument proportional to
$\sqrt{D/|w|}$. Considering the limit of large arguments of
this function \cite{Erdelyi}, we find
\begin{equation}
\label{smallwasy}
p(w;t) \sim
c_{1}(t) \left(\frac{|w|}{D}\right)^{\nu-\frac{1}{2}}
{\rm e}^{- c_{2}(t)\frac{D}{|w|}}
\,\,, \qquad w \to 0^{-}\,\,,
\end{equation}
where
\begin{equation}
\fl
c_{2}(t) = \left(\frac{\nu+1}{\sqrt{2}}\ln\left( 1+t\right) \right)^{2}\,\,,
\qquad
c_{1}(t) = \frac{1}{D}\sqrt{\frac{8}{\pi}} \left(1+t\right)^{\frac{3}{2}}
\!
\left(\frac{2\left(1+t\right)^{\frac{3}{2}}}
{(\nu+1)\ln(1+t)} \right)^{\! \nu}\,\,.
\label{c1c2}
\end{equation}
For any $g$ and any $t$, the PDF almost vanishes in an interval
$(w_{c}(t),0)$, where its width $|w_{c}(t)|$ is controlled by the
``damping constant'' $c_{2}(t)$.  The width increases both with time
and the strength of the logarithmic potential. This can be understood
from the fact that any trajectory yielding a small (absolute) value of
the work, must necessarily depart from a position close to the origin
and remain in its vicinity during the whole time interval $[0,t]$. The
probabilistic weight of such trajectories decreases with both $t$ and
$g$.

\begin{figure}[t]
\begin{center}
\includegraphics[width=1\linewidth]{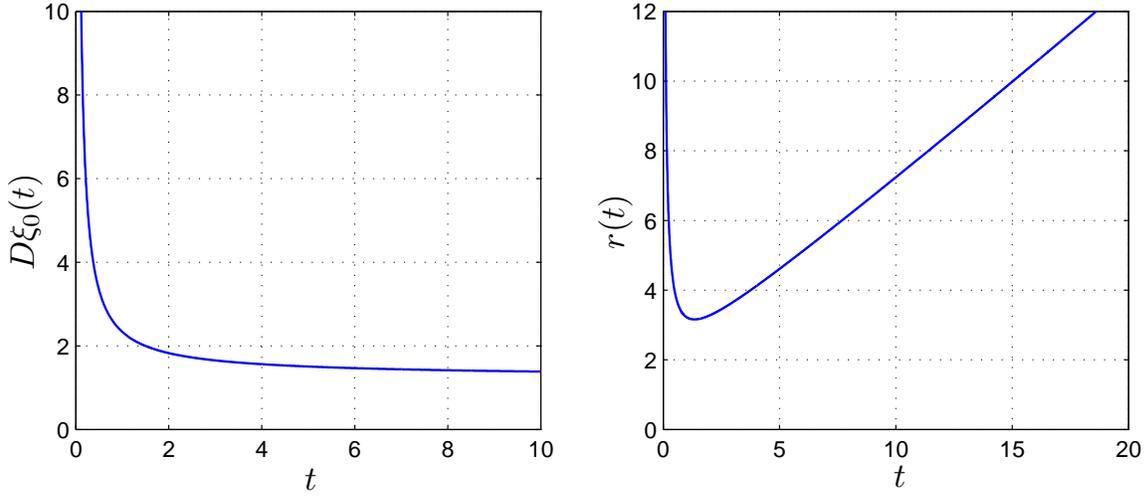}
\caption{Positions of the singularity in the expression
  (\ref{vicinityxi}) (left panel), and the prefactor (\ref{r}) (right
  panel) as functions of time. These functions control, through
  Eq.~(\ref{bigwasy1}), the large $|w|$ asymptotics of the work
  PDF. In the right panel we have taken $D=1$ and $g=1$.}
\label{fig1}
\end{center}
\end{figure}

The asymptotics of the work PDF $p(w;t)$ for $w\!  \to\! - \infty$ (at
fixed $t$) is determined by the expansion of the characteristic
function $\Phi(\xi;t)$ at that $\xi_{0}(t)$, which
gives the singularity of $\Phi(\xi;t)$ lying closest to its domain of
analyticity \cite{Doetsch}. To find $\xi_{0}(t)$, we numerically
solved the transcendental equation $1/\Phi(\xi_{0}(t);t)=0$. In
the vicinity of the singularity,
\begin{equation}
\label{vicinityxi}
\Phi(\xi;t) \sim  r(t)
\left(-\frac{1}{D(\xi - \xi_{0}(t))} \right)^{\nu + 1}\,,
\quad\xi \to \xi_{0}(t)\,\,,
\end{equation}
with
\begin{equation}
\label{r}
r(t)= (-D)^{\nu+1}\!\! \lim_{\xi \to \xi_{0}(t)}
\left(\xi - \xi_{0}(t)\right)^{\nu+1}\Phi(\xi;t)\,\,.
\end{equation}
From this result we obtain \cite{Doetsch}
\begin{equation}
\label{bigwasy1}
p(w;t) \sim \frac{1}{D}\,\frac{r(t)}{\Gamma(\nu+1)}
\left( \frac{|w|}{D} \right)^{\nu} 
{\rm e}^{- D\xi_{0}(t)\frac{|w|}{D}}
\quad,\quad w \to - \infty\,\,,
\end{equation}
where $D\xi_{0}(t)$ is a real, positive, decreasing function of $t$,
cf.\ Fig.~\ref{fig1}. For small $t$, the work PDF is very narrow
($D\xi_{0}(t)$ large). With increasing $t$, the weight of the
trajectories yielding large (absolute) values of the work increases.
This is reflected by the decrease of $D\xi_{0}(t)$.\footnote{For $t=2$
  and $g=0$ we obtain $D\xi_{0}(2) \dot{=} 1.827$,
  $r(2)/\Gamma(1/2)\dot{=}1.021$, which is in perfect agreement with
  Eq.\ (5.30) in \cite{NickelsenEngel}.} Contrary to the function
$c_{2}(t)$ in (\ref{smallwasy}), $D\xi_{0}(t)$ does not depend on the
strength of the logarithmic potential. The parameter $g$ enters only
the pre-exponential factor in Eq.~(\ref{bigwasy1}).

\begin{figure}[t!]
\begin{center}
\includegraphics[width=1.0\linewidth]{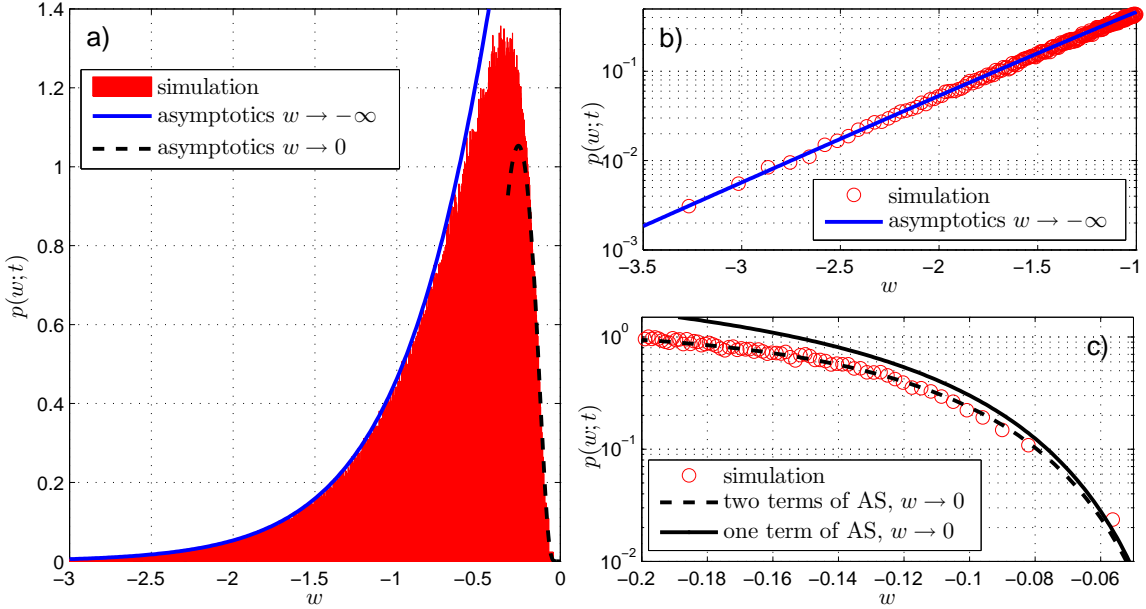}
\caption{a) Simulated work PDF in comparison with the asymptotic
  behavior predicted by Eq.~(\ref{bigwasy1}) ($|w|$ large, solid line)
  and Eq.~(\ref{smallWasy2}) ($|w|$ small, dashed
  line) for parameters $g=1.5$, $D=1$ and $t=1$. In the simulations
  $10^{6}$ trajectories were generated with a time step $\Delta
  t=0.001$ (adapted when the particle is near the origin, see text).
  b) Semi-logarithmic plot of simulated $p(w;t)$ vs.\ $w$ (circles),
  demonstrating the agreement with Eq.~(\ref{bigwasy1}) (solid line)
  for large $|w|$. c) Semi-logarithmic plot of simulated $p(w;t)$ vs.\
  $w$ (circles) in comparison with the first leading term of the
  asymptotic expansion for small $|w|$ (Eq.~(\ref{smallwasy}), solid
  line), and when including the second leading term according to Eq.\
  (\ref{smallWasy2}) (dashed line). }
\label{fig2}
\end{center}
\end{figure}

In order to verify the exact asymptotic expansions of the work PDF, we
have performed extensive Langevin dynamics simulations using the Heun
algorithm \cite{Kloeden} for several sets of parameters and different
time intervals. A typical PDF together with the predictions for its
asymptotic behavior according to Eqs.~(\ref{smallwasy}) and
(\ref{bigwasy1}) is shown in Fig.~\ref{fig2}. In order to avoid
nonphysical negative values of the particle position in the numerics,
which can originate from a fixed time discretization, we have
implemented a time-adapted Heun scheme. If a negative (attempted) 
coordinate along
a trajectory is generated, the time step $\Delta t$ is reduced
until the attempted particle position is positive. To allow for a
better comparison of the analytical findings with the simulated data
in Fig.~\ref{fig2} for small $|w|$, we have derived also the second
leading term in the asymptotic expansion for $|w|\to0$.
After somewhat lengthy but straightforward calculation, we obtain
\begin{equation}
\fl 
p(w;t) \sim
c_{1}(t)
\left[
\left(\frac{|w|}{D}\right)^{\nu-\frac{1}{2}}-
\sqrt{\frac{4}{(\nu +1)\ln(1+t)}}\left(\frac{|w|}{D}\right)^{\nu+\frac{1}{2}}
\right]
{\rm e}^{- c_{2}(t)\frac{D}{|w|}}
\,\,, \quad w \to 0^{-}\,\,,
\label{smallWasy2}
\end{equation}
where $c_{1}(t)$ and $c_{2}(t)$ are given in Eq.~(\ref{c1c2}).

\section{Concluding remarks}

Based on a Lie algebraic approach we succeeded to derive
Eq.~(\ref{finalresult}) for the joint PDF of work and position for a
Brownian particle in a time-dependent logarithmic-harmonic potential.
In order to derive explicit results from Eq.~(\ref{finalresult}) for a
given protocol, the Riccati equation (\ref{Riccati}) needs to be
solved. This nonlinear differential equation is equivalent to the
linear second-order differential equation \cite{PolyaninZaitsev}
\begin{equation}
\ddot{y}(t) - 2 k(t) \dot{y}(t) - 2 D \xi \dot{k}(t) y(t) 
= 0\quad , \quad \dot{y}(0) = 0\,\,.
\label{secondODE}
\end{equation}
Specifically, if $y(t)$ solves (\ref{secondODE}), 
then the logarithmic derivative
\begin{equation}
b_{2}(t) = - \frac{2}{D} \frac{\dot{y}(t)}{y(t)} \,\,,
\label{generalb2}
\end{equation}
is the solution of Eq.\ (\ref{Riccati}). Hence the characteristic
function (\ref{CF2}) can be expressed in terms of the function $y(t)$.

The solution of (\ref{secondODE}) for several reasonable driving
protocols, e.g., for $k(t)=k_{0}\exp(\pm\gamma t)$,
$k(t)=k_{0}+k_{1}t$, or $k(t)=k_{1}t^{n}$, can be written in terms of
higher transcendental functions. Corresponding results are quite
involved and will be published elsewhere. Here we have focused on the
simple protocol (\ref{protocol1}) which should exemplify
typical asymptotic features of the work PDF for {\em monotonic\/}
driving. Notice that, if $\dot{k}(t)>0$ and $\xi\to\infty$ along the
real axis, one can use the WKB approximation and derive a generic
expression for $y(t)$ valid for any protocol, i.e., also a generic
approximative expression for the work characteristic function
(\ref{CF2}).

For non-monotonic driving protocols the work can assume any real value. 
Then the work PDF has the support $(-\infty,+\infty)$ and its two-sided Laplace transform will be
analytic within a stripe parallel to the imaginary axis. The $w\to
+\infty$ ($w\to -\infty$) tail of the work PDF is determined by the
singularity, which is closest to the stripe on its left (right)
side. We hence expect an asymptotics
\begin{equation}
  p(w;t)\sim \frac{1}{D}\frac{r_{\pm}(t)}{\Gamma(\nu+1)}
  \left( \frac{|w|}{D} \right)^{\nu} 
{\rm e}^{- D\xi_{\pm}(t) \frac{|w|}{D} }\quad,\quad w \to \pm \infty\,\,,
\end{equation}
where the coefficients $\xi_{\pm}(t)$, $r_{\pm}(t)$, depend on the
driving protocol $k(t)$. Periodic driving protocols play an
important role in the analysis of Brownian motors.  
A deeper analysis of the work PDF for this class of protocols seems to be worthy for further study.

\ack Support of this work by the Ministry of Education of the Czech
Republic (project No.\ MSM 0021620835), by the Grant Agency of the
Charles University (grant No.\ 143610, and grant No.\ 301311), by the
Charles University in Prague (project No.\ SVV-2012-265 301), and by
the Deutsche Akademische Austauschdienst (DAAD, project No.\
MEB101104) is gratefully acknowledged.

\section*{References}

\end{document}